\documentclass[aps,pra,onecolumn,amsmath,amssymb,groupaddresses,superscriptaddress,11pt]{revtex4}
\usepackage{times}
\usepackage{latexsym}
\usepackage{graphicx}
\usepackage{amsmath,amssymb}
\usepackage{verbatim,bbm}
\newtheorem{theorem}{Theorem}

\begin{document}

\title{Quantum data locking for high-rate private communication}

\author{Cosmo Lupo}
\affiliation{Research Laboratory of Electronics, Massachusetts Institute of Technology, Cambridge, MA 02139, USA}

\author{Seth Lloyd}
\affiliation{Research Laboratory of Electronics, Massachusetts Institute of Technology, Cambridge, MA 02139, USA}
\affiliation{Department of Mechanical Engineering, Massachusetts Institute of Technology, Cambridge, MA 02139, USA}

\begin{abstract}
We show that, if the accessible information is used as a security quantifier, 
quantum channels with a certain symmetry can convey private messages at a tremendously 
high rate, as high as less than one bit below the rate of non-private classical communication.
This result is obtained by exploiting the quantum data locking effect.
The price to pay to achieve such a high private communication rate is that accessible
information security is in general not composable. 
However, composable security holds against an eavesdropper who is forced to measure her share
of the quantum system within a finite time after she gets it.
\end{abstract}

\maketitle

\section{Introduction}\label{sec:intro}

One of the most promising contemporary applications of quantum mechanics
is within cryptography, where the laws of quantum physics 
certify the secrecy of a communication protocol.
In quantum key distribution, the communication protocol aims at
establishing a shared key between two legitimate parties, Alice and Bob, 
in such a way that a third party, say Eve, who eavesdrops on and tampers with
the communication line, obtains virtually no information about the key~\cite{QKD}. 
The key itself is generated randomly, possibly to serve as a one-time pad.
On the other hand, in a private communication protocol, the sender, say Alice, 
aims at sending private messages to Bob~\cite{Private}. 
In this case, the content of the messages is under the control of Alice 
and it is not random from her point of view.
Clearly, any private communication protocol can be also used for key distribution.

In this paper we introduce a private communication protocol, based on
the phenomenon of quantum data locking (QDL)~\cite{QDL}, that achieves a private 
communication rate as high as less than one bit below the classical 
capacity for non-private communication.
Our protocol provides a scheme for realizing a {\it quantum enigma machine}, a quantum
optical cipher based on the QDL effect~\cite{QEM}.
It can be implemented experimentally using standard technologies 
routinely applied in quantum key distribution in setups where information
is encoded by single-photon states spread over $d$ optical modes.
The security of our private communication protocol is assessed in terms of
the accessible information criterion,
which is not the standard and widely accepted security criterion in quantum
cryptography. 
A detailed comparison of the two security criteria is given in \cite{PRL,AW}.
This security criterion is in general weaker than the standard security criterion of quantum cryptography.
For this reason, before proceeding with the description of 
the protocol, we make a brief detour to clarify in which context the accessible 
information yields reliable security, as well as to review the phenomenon of QDL.

\subsection{Accessible information security}\label{sec:acc}

Suppose that Alice's messages are generated by a source
described by the random variable $X$, with probability distribution $p_X(x)$, 
and the conditional states obtained by Eve are $\rho_{E|x}$. 
The ensemble state of the joint system of Alice and Eve is hence
given by the density matrix $\rho_{AE} = \sum_x p_X(x) |x\rangle_A\langle x| \otimes \rho_{E|x}$. 
Let us recall that the accessible information is defined as the maximum classical mutual
information between Alice's input and the result of an optimal measurement performed by Eve
on her share of the quantum system.
A local measurement by Eve is a map $\mathcal{M}_E \, : E \to Y$
whose output is the classical variable $Y$. 
Then the accessible information of the state $\rho_{AE}$ reads
\begin{eqnarray}
I_{acc} = \max_{\mathcal{M}_E} I(X;Y)  \, ,
\end{eqnarray}
where $I(X;Y) = H(X) + H(Y) - H(XY)$ is the classical mutual information, 
and $H$ denotes the Shannon entropy.

To assess the security of our protocol, we show that $I_{acc} \sim \epsilon \log{D}$
where $D$ is the dimension of Eve's quantum system, 
and the security parameter $\epsilon$ can be made arbitrarily small under 
suitable conditions. This means that the outputs of any measurement by Eve are arbitrarily
close to being independent of Alice's messages.
When used as a security quantifier, the accessible information suffers from a major
problem: it does not guarantee composable security.
Roughly speaking, composable security means that if two communication protocols are 
secure individually then they remain secure when composed \cite{compo1,compo2}. The fact that the accessible
information does not ensure composability is intimately related to the very effect
of QDL~\cite{Renner,AW}.
However, as discussed in~\cite{PRX,PRL}, the accessible information yields
composable security conditioned on certain physical assumptions.
A physical assumption that guarantees composable security 
is that the eavesdropper is forced to measure her share of the state as soon as she obtains it,
as is the case, for instance, when she does not have access to a quantum memory.
This is a consequence of the fact that the accessible information concerns the output
of Eve's measurements, and not the quantum state itself.
Another assumption that implies composable security is that Eve possesses a quantum memory with finite coherence time.
In the simplest model, Eve either measures her share of the quantum system within a time $\tau$ 
or the quantum memory decoheres and becomes classical. 
Suppose the given communication protocol is used as a subroutine of a larger protocol. 
Composable security is granted if Alice and Bob know the coherence time
of Eve's quantum memory and wait for a time sufficiently longer than $\tau$ before 
proceeding.
Clearly, too large values of $\tau$ would make the protocol impractical. 
However, as discussed in~\cite{PRL}, in a stationary regime the overall asymptotic communication 
rate is independent of $\tau$ and remains finite even in the limit $\tau \to \infty$.

By making assumptions on the technological capabilities of the eavesdropper
we are in fact restricting the class of allowed attacks.
In quantum cryptography one distinguishes three kinds of attacks: individual attacks (where the eavesdropper
applies local measurement to the output of each use of the communication channel);
collective attacks (where the eavesdropper is allowed to store quantum information for an
indeterminate amount of time before applying a collective measurement on the output of multiple
channel uses); and coherent attack (where the eavesdropper is allowed to tamper with the communication 
line in an arbitrary way).
The assumption that the eavesdropper has a quantum memory with finite coherence time defines
a class of attack that lie in between individual and collective attacks. As in individual attacks,
the eavesdropper cannot store quantum information for an arbitrarily long time. However, in our
case we allow the eavesdropper to store quantum information for a finite time and to apply a 
collective measurement on the output of multiple channel uses.

To be fair, our communication protocol is defined under the assumption that the legitimate
receiver Bob is constrained by the same technological limitations as the eavesdropper Eve.

\subsection{Quantum data locking}\label{sec:QDL}

Below we introduce a private communication protocol that is secure according to the
accessible information criterion. Such a protocol is a QDL protocol. 
In a typical QDL protocol, the legitimate parties, Alice and Bob, publicly
agree on a set of $N = MK$ codewords in a high-dimensional quantum system.  
From this set, they then use a short shared secret key of $\log{K}$ bits to select a set
of $M$ codewords that they will use for sending information.
If the eavesdropper does not know the secret key, then the number of bits,
as quantified by the accessible information, that she can obtain about the message 
is essentially equal to zero for certain choices of codewords.
In most of the known QDL protocols
codewords are chosen from different bases, and the secret key identifies the basis 
to which the codewords belong.
Here we apply a random coding approach and assume that all the $N = MK$ codewords
are chosen randomly.

A number of works have been devoted to the role of QDL in physics and information 
theory~\cite{QDL,CMP,bh,Buhrman,Leung,Fawzi,Dupuis}.
However, only recently QDL has been considered in the presence of noise~\cite{QEM,PRX,phase,AW,PRL}.
A formal definition of the {\it locking capacity} of a communication channel has been 
introduced in~\cite{PRX}, as the maximum rate at which information can be reliably and
securely transmitted through a (noisy) quantum channel $\mathcal{N}_{A \to B}$ from Alice to Bob, 
where the security is quantified by the accessible information.
Motivated by QDL protocols, we also allow the assistance of an initial secret key shared
by Alice and Bob. In order for this key to be inexpensive in the asymptotic limit, we further require that
the bits of secret key grow sublinearly with the number of channel uses.

Two notions of locking capacities were defined in~\cite{PRX}: the {\it weak locking} capacity and the
{\it strong locking} capacity.
The weak locking capacity is defined by requiring security against an eavesdropper who measures 
the output of the complementary channel (denoted as $\mathcal{N}_{A \to E}=\tilde{\mathcal{N}}_{A \to B}$) 
of the channel from Alice to Bob \footnote{We recall that the action of a quantum channel $\mathcal{N}_{A \to B}$ can always be represented as
$\mathcal{N}_{A \to B}(\rho) = \mathrm{Tr}_E \left( V \, \rho \otimes \omega_E \, V^\dag \right)$,
where $\omega_E$ is a pure state of the environment $E$, and $V$ is a unitary transformation coupling the
system with the environment. The conjugate channel of $\mathcal{N}_{A \to B}$ is then defined by
$\tilde{\mathcal{N}}_{A \to B}(\rho) = \mathcal{N}_{A \to E}(\rho) = \mathrm{Tr}_B \left( V \, \rho \otimes \omega_E \, V^\dag \right)$.}.
The strong locking capacity is instead defined by requiring 
security against an eavesdropper who is able to measure the very input of the channel.
In general, the weak locking capacity is larger than or at least equal to the strong locking capacity,
as any strong locking protocol also defines a weak locking one.
It is natural to compare the weak locking capacity with the private capacity~\cite{Private}.
Since the latter is defined by the stronger standard security criterion of quantum cryptography, 
it follows that the weak locking capacity is always larger than or at most equal
to the private capacity.
Finally, both locking capacities cannot exceed the classical capacity,
which is the maximum rate of reliable communication allowed by the channel (not requiring any
secrecy)~\cite{HSW}.
As shown in~\cite{AW}, there exist qudit channels with low or even zero private 
capacity whose weak locking capacity is larger than one half of the classical capacity.
In our previous work, we have obtained key generation protocols that achieve a strong locking 
rate just one bit smaller than the classical capacity~\cite{PRL}.

In a cryptographic setting, the notions of strong and weak data locking capacity correspond to different
kinds of attacks by the eavesdropper.
In a strong locking scenario, we are imaging that the eavesdropper can obtain a noiseless version of the input
states sent by Alice. If the strong locking capacity is nonzero, this mean that these messages can remain locked
to Eve. This is something that cannot happen if the standard security criterion is applied.
In a weak locking scenario, we are instead imagining that the eavesdropper has access to the environment of the
channel. This attack is similar to a collective attack. However, as
discussed in the previous section, the weak locking attack lies in between the collective and individual attacks.

The first result we present in this paper is a QDL protocol for the $d$-dimensional noiseless channel,
see section~\ref{sec:noiseless}.
The protocol allows QDL (in the strong sense) of the noiseless qudit channel at a 
rate of $\log{d}$ bits per channel use, equal to its classical capacity, and consumes 
secret key at an asymptotic rate of less than $1$ bit per channel use.  
The crucial property of this protocol that distinguishes it from prior work on the topic (e.g. \cite{Fawzi})
is that it employs codewords that are separable among different channel uses.%
This property allows us to generalize the protocol to the case of noisy
memoryless channels and to obtain achievable rates of strong and weak locking for a
physically motivated family of qudit channels, see sections~\ref{sec:strong} and~\ref{sec:weak}.

\section{A protocol for strong locking of a noiseless channel}\label{sec:noiseless}

In this section we define a strong locking protocol for direct communication via
a noiseless qudit channel.
This is an improved version of a similar protocol for quantum key distribution that
we have introduced in~\cite{PRL}.
Sections~\ref{sec:pre},~\ref{sec:Eve},~\ref{sec:enet}, and~\ref{sec:impro} present the proof of our main results.
Applications to (weak and strong) locking of 
noisy memoryless channels are then presented in sections~\ref{sec:strong} and~\ref{sec:weak}.

To encode $M$ messages in $n$ qudits, Alice prepares one of the codewords
\begin{equation}
|\psi_c\rangle = \bigotimes_{j=1}^n |x_{j,c}\rangle \, ,
\end{equation}
for $c=1,2,\dots,M$, where the vectors $|x_{j,c}\rangle$ are independently
sampled from an ensemble of qudit states $\{ p(x), |x\rangle \}$ \footnote{We remark that the vectors $|x\rangle$ may not be orthogonal. In general, one could also
replace them with mixed states.
}.
Alice and Bob publicly agree on a set of $K$ $n$-qudit local unitaries
\begin{equation}
U^{(s)} = \bigotimes_{j=1}^n U_j^{(s)}  \, ,
\end{equation}
for $s=1,2,\dots,K$.
According to the value of the secret key,
Alice applies the unitary transformation $U^{(s)}$ to scramble the $n$-qudit codewords, obtaining
\begin{equation}
|\psi_{c}^{(s)}\rangle = U^{(s)} |\psi_{c}\rangle = \bigotimes_{j=1}^n U_j^{(s)} |x_{j,c}\rangle \, .
\end{equation}

In the strong locking scenario, we assume that Eve intercepts the whole train of qudit systems and measures them.
Since Eve does not have access to the secret key, we have to compute the accessible information of the 
state
\begin{equation}\label{AEstate}
\rho_{AE} = \sum_{c=1}^M p(c) | c \rangle \langle c | \otimes \sum_{s=1}^K \frac{1}{K} |\psi_{c}^{(s)}\rangle \langle \psi_{c}^{(s)} |\, ,
\end{equation}
where $\{ | c \rangle \}_{c=1,\dots M}$ is an orthonormal basis for an auxiliary dummy quantum system 
associated to Alice and $p(c)$ is the probability of the codeword $|\psi_{c}\rangle$.
For the sake of simplicity here we assume that all the messages have equal probability, that is, $p(c) = 1/M$
(the case of non-uniform distribution has been considered in~\cite{Fawzi,Dupuis}).
One can upper bound the accessible information as follows (see appendix \ref{app:UB}):
\begin{equation}\label{Iaccn1}
I_{acc} \leq \log{M} - \frac{d^{n}}{M} \, \min_{|\phi\rangle} \left\{  H[Q(\phi)] - \eta\left[ \sum_{c=1}^M Q_c(\phi)\right] \right\} \, ,
\end{equation}
where 
\begin{equation}
Q_c(\phi) = \frac{1}{K} \sum_{s=1}^K |\langle \phi | \psi_{c}^{(s)} \rangle|^2 \, ,
\end{equation}
\begin{equation}
H[Q(\phi)] = - \sum_{c=1}^M Q_c(\phi) \log{Q_c(\phi)} \, ,
\end{equation}
$\eta(x) = - x \log{x}$, and the minimization is over all $n$-qudit unit vectors $|\phi\rangle$.

In the following sections~\ref{sec:pre},~\ref{sec:Eve},~\ref{sec:enet}, and~\ref{sec:impro}, we 
show that there exist choices of the unitaries $\{ U^{(s)} \}_{s=1,\dots,K}$
such that
\begin{equation}
I_{acc} = O \left( \epsilon \log{d^n} \right) \, ,
\end{equation}
provided that
\begin{equation}\label{K-bound-imp-preview}
K > \max\left\{ 2 \gamma^n \,  \left( \frac{1}{\epsilon^2} \ln{M} + \frac{2}{\epsilon^3} \ln{\frac{5}{\epsilon}} \right) , 
\frac{d^n}{M} \, \frac{4 \ln{2} \ln{d^{n}}}{\epsilon^2 } \right\} \, ,
\end{equation}
with
\begin{equation}
\gamma = \frac{2 d}{d+1} \, ,
\end{equation}

In particular, if we put $\epsilon = 2^{-n^s}$ in~(\ref{K-bound-imp-preview}) with $s \in (0,1)$, Eve's accessible information
will be exponentially small in $n$, with an asymptotic secret key consumption rate (in bits per channel use) equal to
\begin{eqnarray}
k & = & \lim_{n\to\infty} \frac{\log{K}}{n} \\
& = & \max\left\{ \log{\gamma} , \log{d} - \lim_{n\to\infty} \frac{\log{M}}{n} \right\} \\
& = & \max\left\{ 1 - \log{\left(1+\frac{1}{d}\right)} , \log{d} - R \right\} \, ,
\end{eqnarray}
where $R = \lim_{n\to\infty} \frac{\log{M}}{n}$.

To show that, we make use of a random coding argument based on random choices of both the
codewords and the data locking unitaries.
In particular, each of the unitaries $U_j^{(s)}$ is generated independently and randomly 
by sampling from the uniform Haar distribution of $d$-dimensional unitaries \cite{NOTAgamma}.

For the case of a noiseless channel, since Bob knows the unitary $U^{(s)}$ chosen by Alice, 
he can simply apply the inverse transformation ${U^{(s)}}^{-1}$ and then perform an optimal 
measurement to discriminate between the codewords \cite{nota-opt}. 
We consider random codewords generated by sampling independently and identically each of 
the qudit state $|x_{j,c}\rangle$ from a given ensemble of input states.
It is well known that in such a setting Bob can decode reliably in the limit 
$n \to \infty$ if $M < \epsilon d^n$, with $\epsilon$ vanishing in the limit $n \to \infty$ \cite{Wilde}.
For instance, putting $\epsilon = 2^{-n^s}$ for $s<1$ one obtains an asymptotic rate
of communication of $R = \lim_{n\to\infty} \frac{1}{n} \log{M} = \log{d}$ bits per channel use, 
with a secret key consumption rate of less than $1$ bit per channel use.

\subsection{Preliminary results}\label{sec:pre}

To characterize our QDL protocol we will make use of two concentration inequalities.
The first one is the tail bound~\cite{AM}:
\begin{theorem}\label{Maurer}
Let $\{ X_t \}_{t=1,\dots,T}$ be $T$ i.i.d.\ non-negative real-valued random variables, with $X_t \sim X$ and 
finite first and second moments, $\mathbb{E}[X],\mathbb{E}[X^2] < \infty$.
Then, for any $\tau > 0$ we have that
$$
Pr\left\{ \frac{1}{T}\sum_{t=1}^T X_t < \mathbb{E}[X] - \tau \right\} \leq \exp{\left( - \frac{T\tau^2}{2\mathbb{E}[X^2]} \right)} \, .
$$
\end{theorem}
($Pr\{ x \}$ denotes the probability that the proposition $x$ is true.)

The second one is the operator Chernoff bound~\cite{Chernoff}:
\begin{theorem}\label{Chernoff}
Let $\{ X_t \}_{t=1,\dots,T}$ be $T$ i.i.d.\ random variables taking values in the algebra of hermitian operators in dimension $D$,
with $0 \leq X_t \leq \mathbb{I}$ and $\mathbb{E}[X_t] = \mu \mathbb{I}$ ($\mathbb{I}$ is the identity operator).
Then, for any $\tau > 0$ and for $(1+\tau)\mu \leq 1$ we have that
$$
Pr\left\{ \frac{1}{T}\sum_{t=1}^T X_t \not\leq (1+\tau)\mu\mathbb{I} \right\} \leq D \, \exp{\left( - \frac{T\tau^2\mu}{4\ln{2}} \right)} \, ,
$$
and
$$
Pr\left\{ \frac{1}{T}\sum_{t=1}^T X_t \not\geq (1-\tau)\mu\mathbb{I} \right\} \leq D \, \exp{\left( - \frac{T\tau^2\mu}{4\ln{2}} \right)} \, .
$$
\end{theorem}


For any given $d^n$-dimensional unit vector $|\phi\rangle$ and codeword $|\psi_{c}^{(s)}\rangle$, 
we define the quantity
\begin{equation}
q_{c}^{(s)}(\phi) = |\langle \phi | \psi_{c}^{(s)} \rangle|^2 = |\langle \phi | U^{(s)} \psi_{c} \rangle|^2 \, .
\end{equation}
Clearly, the latter is a random variable if the 
unitary $U^{(s)}$ and/or the codeword $c$ are chosen randomly.
%
To apply Theorems~\ref{Maurer} and~\ref{Chernoff}, we compute the first and second moments of $q_{c}^{(s)}(\phi)$,
for given $|\phi\rangle$ and $c$, with respect to the i.i.d.\ random locking unitaries.
We obtain (see appendix \ref{app:moments})
\begin{equation}\label{1stm}
\mathbb{E}_U[q_{c}^{(s)}(\phi)] = \frac{1}{d^n} \, , 
\end{equation}
and
\begin{equation}\label{moment2}
\mathbb{E}_U[q_{c}^{(s)}(\phi)^2] \leq \frac{\gamma^n}{d^{2n}} \, ,
\end{equation}
with 
\begin{equation}
\gamma = \frac{2 d}{d+1} \, .
\end{equation}


For any given $|\phi\rangle$ and $c$, we also consider the quantity
\begin{equation}
Q_c(\phi) = \frac{1}{K} \sum_{s=1}^K q_{c}^{(s)}(\phi) \, .
\end{equation}

We now derive several concentration inequalities by applying Theorems~\ref{Maurer} and~\ref{Chernoff}:

\begin{itemize}

\item
Applying Maurer's tail bound (Theorem~\ref{Maurer}), we obtain that for any given $|\phi\rangle$ and $c$
\begin{equation}
Pr\left\{ Q_c(\phi) < \frac{1-\epsilon}{d^n} \right\} \leq \exp{\left( - \frac{K \epsilon^2}{2\gamma^n} \right)} \, .
\end{equation}

We then use this inequality to bound the probability that there exist $\ell$ codewords such that
$Q_c(\phi) < \frac{1-\epsilon}{d^n}$.
Applying the union bound we obtain
\begin{eqnarray}
Pr\left\{ \exists c_1,\dots,c_\ell \, \, | \, \, \forall i \, \, Q_{c_i}(\phi) < \frac{1-\epsilon}{d^n} \right\}
& \leq & { M \choose \ell} \left( Pr\left\{ Q_c(\phi) < \frac{1-\epsilon}{d^n} \right\} \right)^\ell \\
& \leq & { M \choose \ell} \exp{\left( - \frac{\ell K \epsilon^2}{2\gamma^n}\right)} \\
& \leq & M^\ell \,  \exp{\left( - \frac{\ell K \epsilon^2}{2\gamma^n}\right)} \\
& = & \exp{\left( \ell \ln{M} - \frac{\ell K \epsilon^2}{2\gamma^n}\right)} \, . \label{ell-values}
\end{eqnarray}

\item
Let us consider the operators $|\psi_{c}^{(s)}\rangle\langle\psi_{c}^{(s)}|$ and
apply the operator Chernoff bound (Theorem~\ref{Chernoff}).
Notice that equation~(\ref{1stm}) implies
\begin{equation}
\mathbb{E}_U[|\psi_{c}^{(s)}\rangle\langle\psi_{c}^{(s)}|] = \frac{\mathbb{I}}{d^n} \, .
\end{equation}
Putting $\mu = 1/d^{n}$ and $(1+\tau)\mu = (1-\delta)$, the operator Chernoff bound implies that for any given $c$
\begin{eqnarray}
Pr\left\{ \frac{1}{K} \sum_{s=1}^K |\psi_{c}^{(s)}\rangle\langle\psi_{c}^{(s)}| \not\leq 
(1-\delta) \mathbb{I} \right\} & \leq & d^n \exp{\left( -\frac{K (d^n(1-\delta)-1)^2}{d^n 4 \ln{2}} \right)} \\
& = & d^n \exp{\left( -\frac{K d^n (1-\delta-1/d^n)^2}{4 \ln{2}} \right)} \, .
\end{eqnarray}
This in turn implies
\begin{equation}
Pr\left\{ \max_{|\phi\rangle} Q_{c}(\phi) > 1-\delta \right\} \leq d^n \exp{\left( -\frac{K d^n (1-\delta-1/d^n)^2}{4 \ln{2}} \right)} \, .
\end{equation}

We then bound the probability that there exists a codeword $c$ and a
vector $|\phi\rangle$ such that $Q_c(\phi) > 1-\delta$.
Applying the union bound we obtain 
\begin{eqnarray}
Pr\left\{ \max_{|\phi\rangle,c} Q_c(\phi) > 1-\delta \right\} 
& \leq & M Pr\left\{ \max_{|\phi\rangle} Q_c(\phi) > 1 - \delta \right\} \\
& \leq & M d^n \exp{\left( -\frac{K d^n (1-\delta-1/d^n)^2}{4 \ln{2}} \right)} \\
& \leq & \exp{\left( \ln{M d^{n}}  -\frac{K d^n (1-\delta-1/d^n)^2}{4 \ln{2}} \right)} \, . \label{Plarge}
\end{eqnarray}

\item
Finally, we consider random choices of the codewords $c$ and apply the Chernoff bound with $\tau = \epsilon$. 
We then obtain
\begin{equation}
Pr\left\{ \max_{|\phi\rangle} \sum_{c=1}^M Q_c(\phi) \in \left[ (1-\epsilon) \frac{M}{d^{n}} , (1+\epsilon) \frac{M}{d^{n}} \right] \right\} 
\geq 1 - 2 \exp{\left( \ln{d^{n}}-\frac{K \frac{M}{d^{n}} \, \epsilon^2 }{4 \ln{2}} \right)} \, . \label{sum1}
\end{equation}

\end{itemize}

\subsection{Eve's accessible information}\label{sec:Eve}

Let Eve intercept and measure the train of $n$ qudits sent by Alice. 
We now show that, for $n$ large enough, a random choice of the unitaries $U_j^{(s)}$'s guarantees, up to
an arbitrarily small probability, that Eve's accessible information is negligibly small.

We consider a random choice of the codeword $|\psi_c\rangle$.
From equation~(\ref{sum1}), we have that
for all $|\phi\rangle$, $\sum_{c=1}^M Q_c(\phi) \in \left[ (1-\epsilon) \frac{M}{d^{n}} , (1+\epsilon) \frac{M}{d^{n}} \right]$
up to a probability which is bounded away from $1$ provided
\begin{equation}\label{extra-cond}
K > \frac{d^n}{M} \, \frac{4 \ln{2} \ln{d^{n}}}{\epsilon^2 }  \, .
\end{equation}
This yields
\begin{equation}
\frac{d^{n}}{M} \max_{|\phi\rangle} \eta\left[\sum_{c=1}^M Q_c(\phi)\right] 
< \max\left\{ (1-\epsilon) \log\frac{d^{n}}{M} , (1+\epsilon) \log\frac{d^{n}}{M} \right\} \, .
\end{equation}
which in turn implies that, for $K$ large enough, equation~(\ref{Iaccn1}) is 
upper bounded by the following, up to a negligibly small probability,
\begin{eqnarray}\label{Iaccn2-1}
I_{acc} \leq 
\left\{
\begin{array}{lcc}
(1+\epsilon) \log{d^{n}} - \epsilon \log{M} + \eta(1+\epsilon) - \frac{d^{n}}{M} \, \min_{|\phi\rangle} H[Q(\phi)] \, , & \, \, \mbox{for} & M < d^n \, , \\
(1-\epsilon) \log{d^{n}} + \epsilon \log{M} + \eta(1-\epsilon) - \frac{d^{n}}{M} \, \min_{|\phi\rangle} H[Q(\phi)] \, , & \, \, \mbox{for} & M > d^n \, .
\end{array}
\right.
\end{eqnarray}

According to the latter expressions, an upper bound on the accessible information follows
from a lower bound on the minimum Shannon entropy, $\min_{|\phi\rangle} \, H[Q(\phi)]$.
That is, to prove that $I_{acc} \lesssim \epsilon\log{d^n}$, 
we need to show that $\frac{d^{n}}{M} \min_{|\phi\rangle} \, H[Q(\phi)] \gtrsim \left( 1 - \epsilon \right) \log{d^n}$.
To do that, for any $\epsilon > 0$ and $d^n$ and $K$ large enough we bound the probability that 
\begin{equation}
- Q_c(\phi) \log{Q_c(\phi)} < \eta\left( \frac{1-\epsilon}{d^n} \right) \, .
\end{equation}
This corresponds to bounding the probability that either 
$Q_c(\phi) > \lambda_+  = 1 - \eta\left( \frac{1-\epsilon}{d^n} \right) + O\left(\eta\left( \frac{1-\epsilon}{d^n} \right)\right)$
or
$Q_c(\phi) < \lambda_- = (1-\epsilon)/d^n$.
Notice that for $d^n$ sufficiently large and/or $\epsilon$ sufficiently small we have
$\lambda_+ \geq 1 - 2\eta\left( \frac{1-\epsilon}{d^n} \right)$.


First, we bound the probability that there exists a codeword $c$ and a
vector $|\phi\rangle$ such that $Q_c(\phi) > \lambda_+$.
We apply equation~(\ref{Plarge}) with $\delta = 2\eta\left( \frac{1-\epsilon}{d^n} \right)$
to obtain
\begin{eqnarray}
Pr\left\{ \max_{|\phi\rangle,c} Q_c(\phi) > \lambda_+ \right\} 
& \leq & Pr\left\{ \max_{|\phi\rangle,c} Q_c(\phi) > 1 - 2\eta\left( \frac{1-\epsilon}{d^n} \right) \right\} \\
& \leq & \exp{\left( \ln{M d^{n}} -\frac{K d^n (1-2\eta\left( \frac{1-\epsilon}{d^n} \right)-1/d^n)^2}{4 \ln{2}} \right)} \\
& \leq & \exp{\left( \ln{M d^{n}} -\frac{K d^n (1-4\eta\left( \frac{1-\epsilon}{d^n} \right)-2/d^n)}{4 \ln{2}} \right)} \\
& \leq & \exp{\left( \ln{M d^{n}} -\frac{K d^n (1-6\eta\left( \frac{1-\epsilon}{d^n}\right))}{4 \ln{2}} \right)} =: p_+ \, ,
\end{eqnarray}
where we have also used the fact that $\frac{1}{d^n} < \eta\left( \frac{1-\epsilon}{d^n}\right)$ for $n$ large enough.
This probability vanishes exponentially with $d^{n}$ provided $K$ is not too small,
namely, $K > \frac{\ln{M d^{n}}}{d^n} \frac{4 \ln{2}}{1-6\eta[(1-\epsilon)/d^n]}$.


Second, we bound the probability that there exist $\ell \ll M$ codewords such that
$Q_c(\phi) < \lambda_-$.
We apply equation~(\ref{ell-values}) and obtain
\begin{eqnarray}
Pr\left\{ \exists c_1,\dots,c_\ell \, \, | \, \, \forall i \, \, Q_{c_i}(\phi) < \lambda_- \right\}
& = & Pr\left\{ \exists c_1,\dots,c_\ell \, \, | \, \, \forall i \, \, Q_{c_i}(\phi) < \frac{1-\epsilon}{d^n} \right\} \\
& \leq & \exp{\left( \ell \ln{M} - \frac{\ell K \epsilon^2}{2\gamma^n}\right)} \, .
\end{eqnarray}
Putting $\ell = \epsilon M$ we have
\begin{equation}\label{Psmall}
Pr\left\{ \exists c_1,\dots,c_\ell \, \, | \, \, \forall i \, \, Q_{c_i}(\phi) < \lambda_- \right\} 
\leq \exp{\left[ - M \left( \frac{K \epsilon^3}{2\gamma^n} - \epsilon \ln{M} \right) \right]} =: p_- \, .
\end{equation}
Notice that this probability is also exponentially small in $M$, provided that
$K > 2\gamma^n \epsilon^{-2} \ln{M}$.

Inequality~(\ref{Plarge}) implies that, with probability at least equal to $1-p_+$,
all the $Q_c(\phi)$'s are larger than $\lambda_+$.
Also, according to equation~(\ref{Psmall}), for a given $|\phi\rangle$ there exist, with probability greater than $1-p_-$,
at least $M-\ell = (1-\epsilon) M$ values of $c$ such that $Q_c(\phi) > \lambda_-$. 
Putting these results together we obtain that for any given $|\phi\rangle$
\begin{eqnarray}
H[Q(\phi)] & > & - M (1-\epsilon) \left( \frac{1-\epsilon}{d^n} \log{\frac{1-\epsilon}{d^n}} \right) \\
& = & - \frac{M}{d^n} (1-\epsilon)^2 \log{\frac{1-\epsilon}{d^n}} \\
& > & \frac{M}{d^n} (1-2\epsilon) \log{d^n} - \frac{M}{d^n} (1-2\epsilon) \log{(1-\epsilon)} \\
& > & \frac{M}{d^n} (1-2\epsilon) \log{d^n} \, ,
\end{eqnarray}
that is,
\begin{equation}\label{bound-H}
\frac{d^n}{M} H[Q(\phi)] > (1-2\epsilon) \log{d^n} \, ,
\end{equation}
with a probability at least equal to $1 - p_- - p_+$, which is in turn larger than $1 - 2 p_-$
for $M$ large enough.

\subsection{The $\epsilon$-net}\label{sec:enet}

To bound the accessible information in equation~(\ref{Iaccn2-1}) we have to show that a relation
similar to~(\ref{bound-H}) holds for all vectors $|\phi\rangle$. 
To do that we introduce an $\epsilon$-net.
Let us recall that an $\epsilon$-net is a finite set of unit vectors $\mathcal{N}_\epsilon = \{ |\phi_i\rangle \}_i$
in a $D$-dimensional Hilbert space such that for any unit vector $|\phi\rangle$ there exists 
$|\phi_i\rangle \in \mathcal{N}_\epsilon$ for which
\begin{equation}
\| |\phi\rangle\langle\phi| - |\phi_i\rangle\langle\phi_i| \|_1 \leq \epsilon \, .
\end{equation}
As discussed in~\cite{CMP} there exists an $\epsilon$-net with $|\mathcal{N}_\epsilon| \leq (5/\epsilon)^{2D}$.
Below, we first extend the bound~(\ref{bound-H}) to include all the vectors in $\mathcal{N}_\epsilon$, and then, for
$\epsilon$ sufficiently small, to all the manifold of unit vectors.

By applying the union bound we obtain:
\begin{eqnarray}
Pr\left\{ \min_{|\phi_i\rangle \in \mathcal{N}_\epsilon} H[Q(\phi_i)] < (1-2\epsilon) \log{d^n}\right\} 
& \leq & (5/\epsilon)^{2d^n} \, 2 p_- \\ 
& = & 2 (5/\epsilon)^{2d^n} \exp{\left[ - M \left( \frac{K \epsilon^3}{2\gamma^n} - \epsilon \ln{M} \right) \right]} \\
& = & 2 \exp{\left[ - M \left( \frac{K \epsilon^3}{2\gamma^n} - \epsilon \ln{M} - 2 \frac{d^n}{M} \ln{\frac{5}{\epsilon}} \right) \right]} \, .
\end{eqnarray}

Then, we have to replace the minimum over vectors in the $\epsilon$-net with a minimum over all unit vectors. 
An application of the Fannes inequality~\cite{FA} yields (see also~\cite{CMP})
\begin{equation}
\left| \min_{|\phi\rangle} H[Q(\phi)] - \min_{|\phi_i\rangle \in \mathcal{N}_\epsilon} H[Q(\phi_i)] \right| \leq \epsilon \log{d^n} + \eta(\epsilon) \, ,
\end{equation}
which implies
\begin{equation}\label{modify}
Pr\left\{ \min_{|\phi\rangle} H[Q(\phi)] < (1-3\epsilon) \log{d^n} - \eta(\epsilon) \right\} \leq 
2 \exp{\left[ - M \left( \frac{K \epsilon^3}{2\gamma^n} - \epsilon \ln{M} - 2 \frac{d^n}{M} \ln{\frac{5}{\epsilon}} \right) \right]} \, .
\end{equation}
Such a probability is bounded away from one (and goes to zero exponentially in $M$) provided 
\begin{equation}\label{K-bound}
K > 2\gamma^n \,  \left( \frac{1}{\epsilon^2} \ln{M} + \frac{2}{\epsilon^3} \frac{d^n}{M} \ln{\frac{5}{\epsilon}} \right) \, .
\end{equation}

Under this condition for $K$, we finally have the following upper bound for the accessible information
\begin{eqnarray}\label{Iacc-final-nl}
I_{acc} \leq 
\left\{
\begin{array}{lcc}
4 \epsilon \log{d^{n}} - \epsilon \log{M} + \eta(1+\epsilon) + \eta(\epsilon) \, , & \, \, \mbox{for} & M < d^n \, , \\
2 \epsilon \log{d^{n}} + \epsilon \log{M} + \eta(1-\epsilon) + \eta(\epsilon) \, , & \, \, \mbox{for} & M > d^n \, .
\end{array}
\right.
\end{eqnarray}

\subsection{Improving the bound on $K$}\label{sec:impro}

We expect the number of messages to increase exponentially in the number of
channel use, that is, $M \simeq 2^{n R}$. When $2^R < d$, this yields an additional
exponential term, proportional to $d^n/M \simeq (d 2^{-R})^n \gg 1$ on the right hand
side of (\ref{K-bound}).
This term originated from the fact that we are using an $\epsilon$-net on a space
of dimension $d^n$, that contains up to $(5/\epsilon)^{2d^n}$ elements.
We now show that it is sufficient to consider an $\epsilon$-net on a smaller space
of dimension $M$. As a result, we obtain an improved bound on $K$:
\begin{equation}\label{K-bound-imp}
K > 2\gamma^n \,  \left( \frac{1}{\epsilon^2} \ln{M} + \frac{2}{\epsilon^3} \ln{\frac{5}{\epsilon}} \right) \, .
\end{equation}

To show that, we first note that $Q(\phi)$ is indeed a function of
an effective vector $|\tilde\phi\rangle$ with complex components $\tilde\phi_c$, for $c = 1, \dots, M$, 
where
\begin{equation}
|\tilde\phi_c|^2 = \frac{1}{K} \sum_{s=1}^K | \langle \phi | \psi_{c}^{(s)} \rangle | \, .
\end{equation}
For $MK \gg d^n$ the condition (\ref{sum1}) implies that the codewords $ | \psi_{c}^{(s)} \rangle$ fill the whole $d^n$-dimensional
Hilbert space with high probability. This means that we can parameterize any unit vector
$|\phi\rangle$ in terms of the parameters $\tilde\phi_c$ and a set of dummy parameters that
do not affect the value of $Q(\phi)$.

From (\ref{sum1}), we obtain that $\sum_{c=1}^M |\tilde\phi_c|^2 \leq (1+\epsilon)M/d^n \leq 1$, up to small probability.
That is, these parameters define a sphere in $M$ complex dimensions with radius smaller than $1$.
Repeating the same reasoning with an $\epsilon$-net defined on this $M$-dimensional space 
we obtain the bounds (\ref{Iacc-final-nl}) on the accessible information under the tighter condition (\ref{K-bound-imp}) on
the number of key messages.

In conclusion we obtain, from (\ref{Iacc-final-nl}), that 
\begin{equation}
I_{acc} \leq O\left( \epsilon \log{d^n}\right) \, .
\end{equation}
under the condition, from (\ref{K-bound-imp}) and (\ref{extra-cond}),
\begin{equation}\label{K-final}
K \geq \max\left\{ 2\gamma^n \,  \left( \frac{1}{\epsilon^2} \ln{M} + \frac{2}{\epsilon^3} \ln{\frac{5}{\epsilon}} \right) , \frac{d^n}{M} \, \frac{4 \ln{2} \ln{d^{n}}}{\epsilon^2 } \right\} \, .
\end{equation}

\section{Strong locking of a memoryless qudit channel}\label{sec:strong}

The noiseless protocol can be straightforwardly applied for the strong locking of a noisy 
qudit channel $\mathcal{N}_{A \to B}$ connecting Alice to Bob.
The point is that in a strong locking setting we require that the communication is
secure against an eavesdropper having access to the very input of the channel.
In other words, the security of the protocol is independent of how the channel acts
on the input, and hence it applies to the noiseless case as well as the noisy one.
That is, the bound on the accessible information in equation~(\ref{Iacc-final-nl}) and
the condition on the number of key values in equation~(\ref{K-final}) apply for a generic
qudit channel. 

The crucial difference, however, is that the presence of noise reduces the  
rate at which Alice and Bob can reliably communicate classical information.
Let us suppose that, using the codewords described above, Alice and Bob can achieve
a reliable communication rate of $R = \lim_{n\to\infty} \frac{1}{n} \log{M}$ bits per channel use~\cite{DD2007}. 
Then (\ref{K-final}) implies an asymptotic key consumption rate of 
\begin{equation}\label{k-strong}
\lim_{n\to\infty} \frac{\log{K}}{n} = \max\left\{ \log{\gamma} , \log{d} - R \right\}
= \max\left\{ 1 - \log{\left( 1 + \frac{1}{d} \right)} , \log{d} - R \right\} \, .
\end{equation}
Since $R$ cannot exceed $\log{d}$, we obtain an increase in the secret key
consumption rate with respect to the noiseless setting.
We can say that the latter equation represents a trade-off between communication rate and secret key
consumption. In order to achieve strong locking, the secret key consumption rate
should increase to compensate the reduced communication rate.

\section{Weak locking of a memoryless qudit channel}\label{sec:weak}

In the weak locking scenario the eavesdropper has access to the
output of the complementary channel, hence receiving a signal  
distorted by noise. 
One thus expects that the randomness introduced by the noise
contributes to the QDL effect. If this is true, then one can exploit the randomness
due to the noise to reduce the length of the required secret key.
Below we show that this intuition is true by examining a family of channels of a specific
form. We define these channels through their conjugates, which are of the form
\begin{equation}\label{ourch}
\tilde{\mathcal{N}}_{A\to B}(\rho) = \mathcal{N}_{A\to E}(\rho) = p \rho + (1-p) \sigma \, ,
\end{equation}
where $p \in [0,1]$ and $\sigma$ is a given density matrix (notable examples of channels belonging
to this family are the erasure channel and the conjugate of the depolarizing channel).

The results for the noiseless case can be easily applied to these channels.
To do that, it is sufficient to notice that, with probability $p$, the channel $\mathcal{N}_{A\to E}$ is noiseless.
In other words, for $n$ uses of the channel, one expects that the channel $\mathcal{N}_{A\to E}$ will act as an effective
noiseless channel over a fraction of about $pn$ qudits. 
It is sufficient to require that the protocol data locks the information contained in these qudits, since
the remaining $(1-p)n$ qudits do not convey any information at all about the message as the
output is independent of the input.

More formally, upon $n$ uses of the channel Eve receives (with probability arbitrarily close to $1$
for $n$ large enough) no more than $n(p+\delta)$ qudits without any distortion. 
Let us hence consider a given subset of $n(p+\delta)$ qudits and apply the same reasoning of the noiseless channel
given above with $n$ replaced by $n(p+\delta)$.
This yields a bound on Eve's accessible information conditioned on the choice of the subset:
\begin{equation}
Pr\left\{ I_{acc}^{subset} > O \left( \epsilon \log{d^{n(p+\delta)}} \right) \right\} 
\leq 
2 \exp{\left[ - M \left( \frac{K \epsilon^3}{2\gamma^{n(p+\delta)}} - \epsilon \ln{M} - 2\ln{\frac{5}{\epsilon}} \right) \right]}
\end{equation}
(this follows from the bounds in (\ref{modify})).
Finally, we apply the union bound to account for all possible ${ n \choose n(p+\delta)}$ choices of the subset of $n(p+\delta)$ qudits:
\begin{align}
Pr\left\{ I_{acc} > O \left( \epsilon \log{d^{n(p+\delta)}} \right) \right\} 
& \leq 2 { n \choose n(p+\delta)} \exp{\left[ - M \left( \frac{K \epsilon^3}{2\gamma^{n(p+\delta)}} - \epsilon \ln{M} - 2\ln{\frac{5}{\epsilon}} \right) \right]} \\
& \leq 2 n^{n(p+\delta)} \exp{\left[ - M \left( \frac{K \epsilon^3}{2\gamma^{n(p+\delta)}} - \epsilon \ln{M} - 2\ln{\frac{5}{\epsilon}} \right) \right]} \\
& \leq 2 \exp{\left[ n(p+\delta) \ln{n} - M \left( \frac{K \epsilon^3}{2\gamma^{n(p+\delta)}} - \epsilon \ln{M} - 2\ln{\frac{5}{\epsilon}} \right) \right]} \, . \label{Kweak}
\end{align}
This probability goes to zero exponentially in $M$ --- we can always assume that $M = \epsilon 2^{n R}$ where
$R$ is the communication rate --- for $K$ large enough.
From (\ref{Kweak}) and (\ref{extra-cond}), we obtain the following sufficient condition on $K$:
\begin{equation}
K > \max\left\{ 2\gamma^{n(p+\delta)} \,  \left( \frac{1}{\epsilon^2} \ln{M} + \frac{2}{\epsilon^4} \,  \ln{\frac{5}{\epsilon}} \right) , 
\frac{d^{n(p+\delta)}}{M} \, \frac{4 \ln{2} \ln{d^{n(p+\delta)}}}{\epsilon^2 } \right\} \, ,
\end{equation}
which yields an asymptotic secret key consumption rate of (we can assume $\lim_{n\to\infty} \delta = 0$)
\begin{equation}
\lim_{n\to\infty} \frac{\log{K}}{n} = \max\left\{ p \gamma , p \log{d} - R \right\} 
= \max\left\{ p \left[ 1 - \log{\left(1+\frac{1}{d}\right)} \right] , p \log{d} - R \right\} \, .
\end{equation}

This example shows that the presence of noise in the channel to Eve
allows Alice and Bob to consume secret key at a reduced rate, compared to the
strong locking case in (\ref{k-strong}).
We now compute a lower bound on the maximum achievable communication rate
for the class of channels considered here.
To compute $R$, we first write an isometric extension of the channel.
We introduce four quantum systems: systems $1$, $2$ and $3$ are qudits and
system $4$ is a qubit.
In input, system $1$ is assigned to Alice and systems $2$, $3$ and $4$ to Eve.
In output, system $1$ is assigned to Eve and the others to Bob. 
We put
\begin{equation}
U_{1234} = \mathbb{I}_{123} \otimes |0\rangle_4\langle 0|
+ \mathbb{S}_{12} \otimes \mathbb{I}_{3} \otimes |1\rangle_4\langle 1| \, ,
\end{equation}
where $\mathbb{S}_{12}$ is the swap operation between qudits $1$ and $2$.
As initial state of the environment we put 
\begin{equation}
|\varphi_E\rangle = |\varphi\rangle_{23} \otimes \left( \sqrt{p} |0\rangle_4\langle 0| + \sqrt{1-p} |1\rangle_4\langle 1|\right) \, ,
\end{equation}
where $\mathrm{Tr}_3(|\varphi\rangle_{23}\langle\varphi|) = \sigma_2$
(without loss of generality we can also assume $\mathrm{Tr}_3(|\varphi\rangle_{23}\langle\varphi|) = \sigma_3$).

One can easily check that
\begin{equation}
\mathrm{Tr}_{234} \left( U_{1234} \rho_1 \otimes |\varphi_E\rangle_{234}\langle\varphi_E| U_{1234}^\dag \right)
= p \rho_1 + (1-p) \sigma \, .
\end{equation}
Taking the trace over the output systems $1$ we obtain the output of the channel to Bob:
\begin{eqnarray}
\mathrm{Tr}_{1} \left( U_{1234} \rho_1 \otimes |\varphi_E\rangle_{234}\langle\varphi_E| U_{1234}^\dag \right)
& = & p |\varphi\rangle_{23} \langle \varphi| \otimes |0\rangle_4\langle 0| 
+ (1-p) \rho_2 \otimes \sigma_3 \otimes |1\rangle_4\langle 1| \nonumber \\
& + & \sqrt{p(1-p)} \left[ \mathrm{Tr}_1 \left( \mathbb{S}_{12} \rho_1 \otimes|\varphi\rangle_{23}\langle\varphi| \right) \otimes |1\rangle_4\langle 0| + \mathrm{h.c.} \right]
\, .
\end{eqnarray}
We notice that the action on the channel from Alice and Bob depends on $\sigma$ through
the last two terms proportional to $|0\rangle_4 \langle 1|$ and $|1\rangle_4 \langle 0|$.
If we apply a completely dephasing channel on qubit $4$ the channel to Bob
becomes an erasure channel with erasure probability $p$ independently of $\sigma$.
This implies that the classical capacity of the erasure channel is an achievable 
rate for classical communication, hence we can put $R = (1-p) \log{d}$.
Moreover, this bound holds for any choice of the locking unitary, since the erasure channel is 
covariant under unitary transformations.

\subsection{Erasure channel}\label{sec:erasure}

If $\sigma$ is orthogonal to the input space, the channel
in equation~(\ref{ourch}) is a qudit erasure channel with erasure probability $1-p$,
whose complement is an erasure channel with erasure probability $p$.
In this case, the maximum communication rate equals the classical capacity of the
erasure channel, $R = (1-p)\log{d}$, with a secret key consumption rate of  
\begin{equation}
k = \max\left\{ p \left[ 1 - \log{\left( 1 + \frac{1}{d} \right)} \right] , (2p-1) \log{d} \right\}
\end{equation}
bits per channel use.

\subsection{Conjugate of the depolarizing channel}

If $\sigma = \mathbb{I}/d$, the channel in equation~(\ref{ourch}) is a qudit 
depolarizing channel with depolarizing probability $1-p$.
We can rewrite the action of the depolarizing channel as
\begin{equation}
\mathcal{N}_{A\to E}(\rho) = p \rho + \frac{1-p}{d^2} \sum_{a,b=0}^{d-1} X^a Z^b \rho Z^{-b} X^{-a} \, , 
\end{equation}
where $X = \sum_{j=0}^{d-1} |j \oplus 1\rangle \langle j|$ ($\oplus$ denotes summation
modulo $d$) and
$Z = \sum_{j=0}^{d-1} e^{i j 2\pi/d}|j \rangle \langle j|$ are the $d$-dimensional generalization of the Pauli matrices, 
and $\{ |j\rangle \}_{j=0,\dots,d-1}$ is a qudit basis.
This representation of the channel to Eve induces a representation for the isometric extension, 
which is given by the bipartite conditional unitary
\begin{equation}
U = \sum_{a,b=0}^{d-1} |ab\rangle \langle ab| \otimes X^a Z^b \, ,
\end{equation}
where the first system, assigned to Eve's input, is represented by a $d^2$-dimensional Hilbert space
(spanned by the basis vectors $\{ |ab\rangle \}$), and the second is the input qudit system.
As initial state of Eve's system we take
\begin{equation}
|\varphi_E\rangle = \sum_{a,b=0}^{d-1} \sqrt{q_{ab}} \, |ab\rangle \, ,
\end{equation}
where $q_{00} = p + (1-p)/d^2$ and $q_{ab} = (1-p)/d^2$ for $ab \neq 00$.
Taking the partial trace over Eve's output system, we finally obtain the following expression for the channel to Bob:
\begin{equation}
\mathcal{N}_{A \to B}(\rho) = \sum_{a,b,a',b'=0}^{d-1} \sqrt{q_{ab} q_{a'b'}} \, \mathrm{Tr}\left( X^a Z^b \rho Z^{-b'} X^{-a'} \right) \, |ab\rangle\langle a'b'|  \, .
\end{equation}
A straightforward calculation yields that the maximum
achievable rate using our ensemble of input states is
\begin{equation} \label{dep-rate}
R = f(p,d^2) - f(p,d) \, ,
\end{equation}
where
\begin{equation}\label{f-fun}
f(p,D) = - \left( p + \frac{1-p}{D} \right)\log{\left( p + \frac{1-p}{D} \right)} - \left( D - 1 \right) \frac{1-p}{D} \log{\left(\frac{1-p}{D}\right)} \, .
\end{equation}
As in the case of the erasure channel, this rate is independent of the choice of 
the locking unitary.

In conclusion, equation~(\ref{dep-rate}) gives the maximum reliable communication rate
from Alice to Bob. 
The secret key consumption rate is hence equal to
\begin{equation}
k = \max{\left\{ p \left[ 1 - \log{\left( 1+\frac{1}{d} \right)} \right] , p \log{d} - f(p,d^2) + f(p,d) \right\}} \, .
\end{equation}

\section{Quantum bootstrapping}

Our QDL protocols yield achievable rates of strong and weak locking through a noisy qudit channel.
Consider the case of weak locking at a rate $R$ with a secret consumption rate of $k$ bits per channel use.
If the channel $\mathcal{N}_{A \to B}$ has non-zero private capacity, one can first use 
the channel to establish a private key between the two legitimate parties, then use such 
a key to lock the subsequent uses of the channel. 
In this way Alice and Bob will achieve a weak locking rate of 
\begin{equation}
R_\mathrm{wl} = \frac{R}{1+k/P} \, , 
\end{equation}
where $P$ is the private capacity of the communication channel~\cite{Private}.
It follows that any channel with non-zero private capacity allows us to lock 
data in the weak locking sense.
For qudit channels having $R$ and $P$ large enough, $R_\mathrm{wl}$ 
can be much larger than the private capacity and arbitrarily close to the classical capacity.

Alternatively, for generic channels, including those with zero private capacity, one can 
define a QDL protocol under the assumption that Alice and Bob know an upper bound $\tau$
on the coherence time of Eve's quantum memory.
%
For $R > k$, a weak locking protocol is then defined according to the following procedure:
\begin{enumerate}

\item
Alice and Bob initially share a secret key of $n k$ bits;

\item
They use the secret key to send about $n R$ bits of locked 
information through $n$ uses of the qudit channel;

\item
They wait a time $\tau$ sufficiently long to guarantee that Eve's quantum 
memory decoheres. After such a time the locked information Alice has sent
to Bob can be considered secure in the composable sense (see Sec.~\ref{sec:intro} and Ref.~\cite{PRL});

\item
If $R > k$, Alice and Bob recycle $n k$ of the $n R$ bits as a secret key 
for the next round of the communication protocol;

\item
They repeat the above procedure for $n'$ times.

\end{enumerate}

(We remark that Bob does not need to store quantum information for a
time longer than $\tau$. Indeed, he needs to store quantum information only for
the time necessary to send $n R$ bits along the channel from Alice to Bob.
In other words the protocol does not require the legitimate Bob to have better
technology than the eavesdropper Eve.)

Using this bootstrap technique, Alice and Bob will asymptotically achieve a weak locking
rate of (for $R \geq k$)
\begin{equation}
R_\mathrm{wl} = R - k 
\end{equation}
bits per channel use, with a secret key consumption rate of $k/n'$ bits 
that goes to zero in the limit $n' \to \infty$.
While the rate per channel use is finite and independent of $\tau$, one may object
that the communication rate {\it per second} will become arbitrarily small if $\tau$ is large 
enough. To solve this problem, Alice and Bob can run two or more independent instances of the protocol
in parallel (each using an independent secret key) taking advantage of the dead times between one 
protocol and the other.
It follows that the communication rate per second remains 
finite and independent of $\tau$ even in the limit of $\tau \to \infty$.
(Clearly, this procedure becomes impractical if $\tau$ is too large.)

\begin{figure}
\centering
\includegraphics[width=0.45\textwidth]{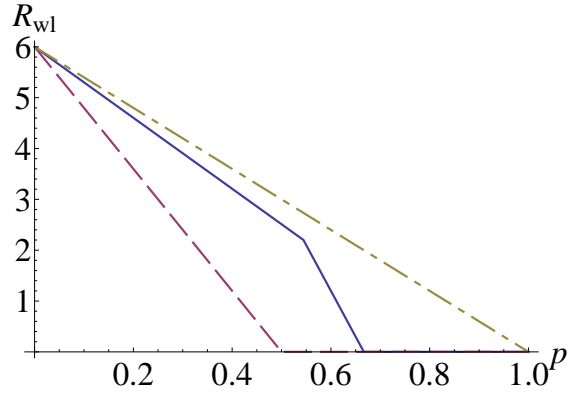}
\caption{Comparison of several communication rates (in bits per channel use) 
for the qudit erasure channel, with $d=64$ and $p \in [0,1]$.
Weak locking rate (solid line);
private capacity (dashed line);
classical capacity (dot-dashed line).
} \label{fig:erasure}
\end{figure}

\begin{figure}
\centering
\includegraphics[width=0.45\textwidth]{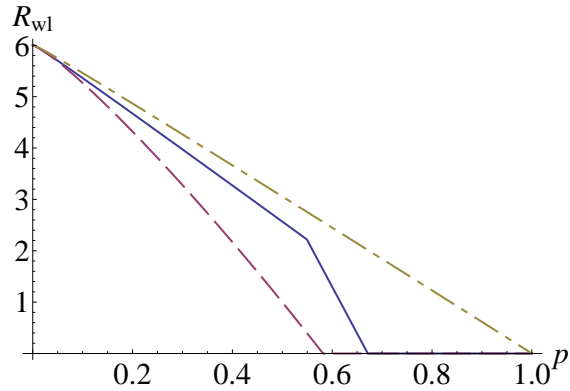}
\caption{Comparison of several communication rates (in bits per channel use) 
for the conjugate of the depolarizing channel, with $d=64$ and $p \in [0,1]$.
Weak locking rate (solid line);
hashing bound for private communication (dashed line);
classical capacity (dot-dashed line).
} \label{fig:depolarizing}
\end{figure}

For the qudit erasure channel, the procedure described above achieves a weak locking rate (in bits per channel use) of 
\begin{equation}
R_\mathrm{wl} = (1-p)\log{d} - \max\left\{ p \left[ 1 - \log{\left( 1 + \frac{1}{d} \right)} \right] , (2p-1)\log{d} \right\}   
\end{equation}
Similarly, for the conjugate of the qudit depolarizing channel we obtain
\begin{equation}
R_\mathrm{wl} = f(p,d^2) - f(p,d) - \max\left\{ p \left[ 1 - \log{\left( 1 + \frac{1}{d} \right)} \right] , p\log{d} - f(p,d^2) + f(p,d) \right\} \, ,
\end{equation}
where $f(p,d)$ and $f(p,d^2)$ are as in equation~(\ref{f-fun}).
Figure~\ref{fig:erasure} shows the weak locking rate of the qudit erasure channel 
compared with the classical capacity~\cite{HSW} and the private capacity~\cite{Private}.
Figure~\ref{fig:depolarizing} shows the weak locking rate of the conjugate of the qudit depolarizing
channel, compared with its classical capacity and the Hashing bound for private communication.

The idea of key recycling is not new in quantum cryptography (see, e.g., \cite{BB82}). 
The crucial difference in our approach is that we are assuming the weaker security criterion
expressed in terms of the accessible information. By weakening the notion of security 
we are able to obtain a positive rate of locked communication even if
the channel has zero privacy according to the standard security criterion of
quantum cryptography.

\section{Conclusions}

In conclusion, we have presented protocols that achieve a weak locking rate
as high as less than one bit below the classical capacity for quantum 
channels exhibiting certain symmetry properties. 
These results, together with~\cite{AW,PRL}, further deepen our understanding 
of the QDL effect as well as of the notions of locking capacities recently introduced in~\cite{PRX}.
A few natural questions remain open. It is not clear whether our strong locking
protocol for the noiseless channel is optimal in terms of secret key consumption.
The obtained secret key consumption rate of $\max\{ 1 - \log{(1 + 1/d)} , \log{d} - R \}$ bits per channel use could very well not be
a fundamental limit, but just a consequence of our proof technique.
Also, one would like to find weak locking protocols for general channels beyond
the restricted, yet physically relevant, class of channels considered here.
Finally, since the most important realizations of quantum communication channels
are within continuous-variable, it is urgent to discover QDL protocols
for quantum systems with infinite dimensions.

\vspace{0.5cm}

{\bf Acknowledgment.} We are grateful to Giuseppe Vallone, Fr\'ed\'eric Dupuis, 
Andreas Winter, and Mark M. Wilde for helpful discussions and comments. 
This research was supported by the DARPA Quiness Program through 
U.S.~Army Research Office Grant No.~W31P4Q-12-1-0019.


\appendix

\section{Upper bound on the accessible information }\label{app:UB}

In this appendix we derive an upper bound on the accessible information of the bipartite state
\begin{equation}
\rho_{AE} = \frac{1}{M} \sum_{x=1}^M | x \rangle \langle x | \otimes \sum_{s=1}^K \frac{1}{K} |\psi_{x}^{(s)}\rangle \langle \psi_{x}^{(s)} | \, .
\end{equation}

The accessible information is the maximum classical mutual
information between Alice's input $X$ and the result of an optimal 
measurement performed by Eve on her share of the quantum system.
Such a local measurement is described by a set of POVM elements $\{ \Lambda_y \}_{y}$,
with $\Lambda_y \geq 0$ and $\sum_{y} \Lambda_y = \mathbb{I}$, where
$y$ is the value of the corresponding measurement result. 
The output of the measurement is a random variable $Y$. The conditional probability
distribution of $Y$ given $x$ is
\begin{equation}
p_Y(y|x) = \sum_{s=1}^K \frac{1}{K} \langle \psi_{x}^{(s)} | \Lambda_y |\psi_{x}^{(s)}\rangle \, ,
\end{equation}
and $p_Y(y) = M^{-1} \sum_x p_Y(y|x)$

Then the accessible information of the state $\rho_{AE}$ reads
\begin{eqnarray}
I_{acc} = \max_{ \{ \Lambda_y \} } \, I(X;Y) = \max_{ \{ \Lambda_y \} } \, H(X) + H(Y) - H(XY) \, ,
\end{eqnarray}
where 
\begin{equation}
H(X) = - \sum_x p_X(x) \log{p_X(x)} = \log{M} \, ,
\end{equation}
\begin{equation}
H(Y) = - \sum_y p_Y(y) \log{p_Y(y)} \, ,
\end{equation}
and 
\begin{equation}
H(XY) = - \sum_{x,y} p_Y(y|x)p_X(x) \log{p_Y(y|x)p_X(x)} = - \sum_{x,y} M^{-1} p_Y(y|x) \log{M^{-1} p_Y(y|x)} \, .
\end{equation}

By convexity of mutual information, it is sufficient to restrict to the set 
of rank-one POVM with $\Lambda_y = \mu_y |\phi_y\rangle \langle \phi_y|$,
where the $|\phi_y\rangle$'s are unit vectors and $\mu_y >0$.
The condition $\sum_y \mu_y |\phi_y\rangle \langle \phi_y| = \mathbb{I}$ 
then implies $\sum_y \mu_y/d^n = 1$.
A straightforward calculation yields
\begin{equation}\label{Iacc-exp}
I_{\text{acc}} = \log{M} - 
\min_{ \{ \mu_y|\phi_y\rangle\langle\phi_y| \} } \sum_y \frac{\mu_y}{M} \left\{ H[Q(\phi_y)] - \eta\left[ \sum_{x} Q_x(\phi_y)\right] \right\} \, ,
\end{equation}
where $\eta( \cdot ) = - ( \cdot ) \log{( \cdot )}$,
$Q(\phi_y)$ is the $M$-dimensional real vector of non-negative components
\begin{equation}
Q_x(\phi_y) = \frac{1}{K} \sum_{s=1}^K |\langle \phi_y | \psi_{x}^{(s)} \rangle |^2 \, ,
\end{equation}
and $H[Q(\phi_y)] = - \sum_x Q_x(\phi_y) \log{Q_x(\phi_y)}$.

We now apply a standard convexity argument, first used in~\cite{QDL}. 
To do that, notice that the positive quantities $\mu_y/d^n$ can be interpreted as probability weights. 
An upper bound on the accessible information (\ref{Iacc-exp}) is then obtained by using the fact that
the average cannot exceed the maximum.
This yields
\begin{eqnarray}
I_{\text{acc}} & = & 
\log{M} - \frac{d^n}{M}\min_{ \{ \mu_y|\phi_y\rangle\langle\phi_y| \} } \sum_y \frac{\mu_y}{d^n} \left\{ H[Q(\phi_y)] - \eta\left[ \sum_{x} Q_x(\phi_y)\right] \right\} \\
& \leq & \log{M} - \frac{d^n}{M} \min_{|\phi\rangle} \left\{ H[Q(\phi)] - \eta\left[\sum_x Q_x(\phi)\right] \right\} \, ,
\end{eqnarray}
which is the upper bound in (\ref{Iaccn1}).


%

\section{Calculation of the first and second moment}\label{app:moments}

Here we compute the first and second moment of $q_c(\phi) = | \langle \phi | U | \psi \rangle|^2$
with respect to a random unitary of the form $U = \otimes_{j=1}^n U_j$, where each 
qudit unitary $U_j$ is independently sampled from the uniform distribution induced by the Haar measure 
$d\mu(U_j)$ on the unitary group.

We have
\begin{eqnarray}
\mathbb{E}_U[q_c(\phi)] & = & \mathbb{E}_U[\langle \phi | U |\psi_c \rangle \langle \psi_c | U^\dag | \phi \rangle] \\
& = & \langle \phi | \mathbb{E}_U[ U |\psi_c \rangle \langle \psi_c | U^\dag ] | \phi \rangle \\
& = & \langle \phi | \otimes_{j=1}^n \mathbb{E}_{U_j}[ U_j |x_{j,c} \rangle \langle x_{j,c} | U_j^\dag ] | \phi \rangle \\
& = & \frac{1}{d^n} \, ,
\end{eqnarray}
where we have used
$\mathbb{E}_{U_j}[ U_j |x \rangle \langle x | U_j^\dag ] = \int d\mu(U_j) U_j |x \rangle \langle x| U_j^\dag = \mathbb{I}/d$
for any unit vector $|x \rangle$.

To compute the second moment we first write
\begin{eqnarray}
\mathbb{E}_U[q_c(\phi)^2] & = & \mathbb{E}_U[\langle \phi | U |\psi_c \rangle \langle \psi_c | U^\dag | \phi \rangle
\langle \phi | U |\psi_c \rangle \langle \psi_c | U^\dag | \phi \rangle] \\
& = & \mathbb{E}_U[\langle \phi,\phi | U \otimes U |\psi_c,\psi_c \rangle \langle \psi_c,\psi_c | U^\dag \otimes U^\dag | \phi,\phi \rangle] \\
& = & \langle \phi,\phi | \mathbb{E}_U[ U \otimes U |\psi_c,\psi_c \rangle \langle \psi_c,\psi_c | U^\dag \otimes U^\dag ] | \phi,\phi \rangle \\
& = & \langle \phi,\phi | \otimes_{j=1}^n \mathbb{E}_{U_j}[ U_j \otimes U_j |x_{j,c},x_{j,c} \rangle \langle x_{j,c},x_{j,c} | U_j^\dag \otimes U_j^\dag ] | \phi,\phi \rangle \\
& = & \langle \phi,\phi | \otimes_{j=1}^n 
\int d\mu(U_j) U_j \otimes U_j |x_{j,c},x_{j,c} \rangle \langle x_{j,c},x_{j,c} | U_j^\dag \otimes U_j^\dag
| \phi,\phi \rangle \, ,
\end{eqnarray}
where $|\phi,\phi\rangle = |\phi\rangle \otimes |\phi\rangle$, $|\psi_c,\psi_c\rangle = |\psi_c\rangle \otimes |\psi_c\rangle$
and $|x_{j,c},x_{j,c}\rangle = |x_{j,c}\rangle \otimes |x_{j,c}\rangle$.
We then apply the representation of the twirling operator \cite{WS}
\begin{eqnarray}
\mathcal{T}(\rho) & = & \int d\mu(U) U \otimes U \, \rho \,  U^\dag \otimes U^\dag \\
& = & \mathrm{Tr}(\rho Q_0) \frac{Q_0}{\mathrm{Tr(Q_0)}} + \mathrm{Tr}(\rho Q_1) \frac{Q_1}{\mathrm{Tr(Q_1)}} \, ,
\end{eqnarray}
where
\begin{equation}
Q_\alpha = \frac{\mathbb{I} + (-1)^\alpha S}{2}
\end{equation}
are the projectors on the symmetric ($\alpha=0$) and anti-symmetric ($\alpha=1$) subspaces, 
$\mathbb{I}$ denotes the identity operator, and $S$ is the swap operator ($S |\psi,\psi'\rangle = |\psi',\psi\rangle$).
We then have
\begin{equation}
\mathcal{T}(|x_{j,c},x_{j,c} \rangle \langle x_{j,c},x_{j,c} |) = \frac{Q_0}{\mathrm{Tr}(Q_0)} = \frac{2}{d(d+1)} \, Q_0 \, ,
\end{equation}
which yields
\begin{eqnarray}
\mathbb{E}_U[q_c(\phi)^2] 
& = & \left( \frac{2}{d(d+1)} \right)^n \langle \phi,\phi | Q_0^{\otimes n} | \phi,\phi \rangle \\
& \leq & \left( \frac{2}{d(d+1)} \right)^n \| Q_0^{\otimes n} \|_\infty \\
& = & \left( \frac{2}{d(d+1)} \right)^n \| Q_0 \|_\infty^n \\
& = & \left( \frac{2}{d(d+1)} \right)^n \, .
\end{eqnarray}
Here we have used the fact that 
$\langle \phi,\phi | Q_0^{\otimes n} | \phi,\phi \rangle \leq \| Q_0^{\otimes n} \|_\infty$,
where $\| Q_0^{\otimes n} \|_\infty$ is the operator norm of $Q_0^{\otimes n}$ (namely, the supremum of its eigenvalues), 
and that $\| Q_0^{\otimes n} \|_\infty = \| Q_0 \|_\infty^n = 1$.


\begin{thebibliography}{99}
\bibitem{NOTAgamma}
The value of $\gamma$ depends on the ensemble of unitaries used to scramble
the codewords. This value is obtained if the unitaries are sampled from the
uniform Haar distribution --- see equation (\ref{moment2}) and appendix \ref{app:moments}.
In \cite{PRL,phase} different values of $\gamma$ were obtained by applying other ensembles
of scrambling unitaries.
\bibitem{nota-opt}
By optimal measurement we mean any measurement that achieves the 
Holevo bound as, e.g., the pretty good measurement \cite{HSW}.
\end{thebibliography}
\end{document}